\documentclass{article}
\usepackage[T1]{fontenc}
\usepackage[utf8]{inputenc}
\usepackage{graphicx}
\usepackage{color,soul}
\usepackage{url} 
\usepackage{hyperref}
\usepackage{adjustbox}
\usepackage[margin=3.75cm]{geometry}
\usepackage{caption}
\usepackage{subcaption}
\usepackage{gnuplot-lua-tikz}
\usepackage{tikz}
\usepackage{tikzscale}
\usepackage{pgfplots}
\usepackage{authblk}
\hypersetup{
   colorlinks,
   citecolor=blue,
   filecolor=blue,
   linkcolor=blue,
   urlcolor=blue
}

\usepackage{makeidx}  
\begin{document}

\title{Investigating the Evolvability  of Web Page Load Time}
\author[1]{Brendan Cody-Kenny}
\author[2]{Umberto Manganiello}
\author[2]{John Farrelly}
\author[2]{Adrian Ronayne}
\author[2]{Eoghan Considine}
\author[2]{Thomas McGuire}
\author[1]{Michael O'Neill}
\affil[1]{Natural Computing Research and Applications Group (NCRA), 
  Michael Smurfit Graduate Business School, 
  University College Dublin, 
  Ireland.
}
\affil[2]{Fidelity Investments, Ireland.}

\maketitle              

\begin{abstract}
  Client-side Javascript execution environments (browsers) allow anonymous functions and event-based programming concepts such as callbacks. 
  We investigate whether a mutate-and-test approach can be used to optimise web page load time in these environments. 
  First, we characterise a web page load issue in a benchmark web page and derive performance metrics from page load event traces. We parse Javascript source code to an AST and make changes to method calls which appear in a web page load event trace. We present an operator based solely on code deletion and evaluate an existing ``community-contributed'' performance optimising code transform. 
  By exploring Javascript code changes and exploiting combinations of non-destructive changes, we can optimise page load time by 41\% in our benchmark web page. \\
  
\end{abstract}

\section{Introduction}
Performance characteristics vary across browsers where improvements in one version may degrade performance in another \cite{selakovic2016performance}. Performance characteristics also change frequently as a Javascript engine is subject to re-design. As a result, performance tuning is a never-ending task. Javascript developers optimise code for Javascript engines while Javascript engine developers optimise for how the engine will likely be used.

While a range of work has looked at mutation-based performance~\cite{langdon2015performance,cody2015locogp} and energy improvement~\cite{Bokhari:2017:GI}, no work we are aware of has inspected source code mutation for page load time in the browser. Related work has looked at web service component selection~\cite{chang2005optimizing}, though this targets components which are a higher level granularity of software unit.

In this paper we investigate the base mechanisms needed for a Genetic Programming (GP) code improvement system: fitness measures and operators. We trace page load for a simple benchmark web app and calculate (i) time, (ii) number of events, and (iii) the largest depth of event chains found in the trace. We inspect two operators, one which deletes statements and expressions containing method names found in the trace, and another which transforms loops to more optimal versions. 
To validate these mechanisms, we apply each operator iteratively to all source code locations where they are applicable using a greedy search loop. After applying an operator, if the web page appears as expected, we keep the source code mutation.

Our code deletion operator was able to reduce (i) page load time by 41\%, (ii) total events by 30\% and (iii) event depth by 26\%. While these results are encouraging they are relevant only to our benchmark application, which contains much redundancy by design. 

We cover related work in the following section, with Section~\ref{exptlsetup} providing the experimental setup, the target web application is described in Section~\ref{webapp}. We then outline the observations arising from the experiments in Section~\ref{results} before drawing conclusions and suggesting directions for future work (Section~\ref{conc}).

\section{Related Work}

Previous work on performance improvement has focused on run-time~\cite{langdon2015performance,cody2015locogp} and energy improvement~\cite{Bokhari:2017:GI}.
Program performance improvements frequently result from code deletion~\cite{langdon2013faster}, motivating us to initially investigate a deletion operator.
Repeatedly applying deletion produces a sub-program similar to one which could be found using program slicing techniques~\cite{binkley2004survey,ye2016efficient}. 
Designing operators for performance improvement is an open problem, though it has been proposed that new operators may be derived by mining existing code and code generated during GP runs~\cite{Petke:2017:GI}. Program transforms are also written and released by developers in the spirit of making useful transforms reusable~\cite{jscodemod}, adding another potential source of operators.

Exhaustive mutation has been used to find how robust program functionality is to source code change.
Mutating small programs with fine-grained operators in a relatively statically typed language such as Java appears to result in relatively low mutational robustness of 30\%~\cite{cody2015locogp}, while larger programs in C++ appear to have high mutational robustness of up to 89\%~\cite{langdon:2015:csdc,schulte2014software} and 68\% in more recent work~\cite{langdon2017evolving}. As mutational robustness varies depending on the software under evolution, we currently have relatively few data points to draw comparisons.

When using search algorithms on large programs it is important to focus operators to reduce search space size~\cite{DBLP:conf/gecco/ForrestNWG09,langdon2013faster}. More targeted mutation operators perform program transforms which are highly likely to impact performance while leaving functionality unchanged. 
For example, multiple different list implementations which have the same interface can be evaluated~\cite{basios2017darwinian}.

\section{Experimental Setup}
\label{exptlsetup}
We gather performance measures based on (i) page load time, (ii) number of events during this time as well as (iii) the depth of event chains. We investigate two operators, one written by the authors which deletes code based on method names and one which has been made available as a community contribution. The search loop used simply keeps a code change if the page does not show any error.

\subsection{Metrics for Web Page Load}
We are mainly interested in improving \textbf{page load time} for a web app.
We gather traces of the web app loading via chrome browser's devtools functionality\footnote{https://developer.chrome.com/devtools} (with caching disabled) using a client for NodeJS\footnote{https://github.com/cyrus-and/chrome-remote-interface}. Page load traces list events which can be used to build a call graph for the entire page load process. The elapsed time for all events to complete and the depth of call sequences can be calculated from a page trace. We take the time between the first and last event recorded in the browser to give load time. 

We sum browser \textbf{events} as a pseudo-measure of performance, which is subject to less variation than time-based measures. The number of events in total gives us an idea of how much work is being performed by the browser. We also take a measure of the most deeply nested sequence of event calls, which gives us a rough idea of how interdependent events are. Our intuition is that time can elapse while this call graph is traversed even though more computation could be performed during the same elapsed time if dependent method calls could be rearranged. 

We assume we are beginning with an ``oracle'' web app which is considered \textbf{correct}. 
During page load, we check what elements have been loaded onto the page, and the page is only considered fully loaded when a string appears as part of the Document Object Model (DOM) for the page. We sum the number of pixels by which screenshots differ and use this as our only measure of functional correctness\footnote{http://www.imagemagick.org/Usage/compare/\#statistics} (as this is the only functionality our benchmark provides). We compare two oracle screenshots to get a measure of acceptable variation while still considering the web app correct.  We use a multiple of this acceptable pixel variability measure to get a threshold value, above which the page is considered incorrect, that is, too different to be considered the same as the original. Subsequent screenshots taken of web app variants are compared with the original oracle screenshot.

A screenshot captures the final state of the page after it has been loaded. The screenshot does not capture anything about the underlying state of the web page or the structure of the web page Document Object Model (DOM). As a result, using screenshots to measure correctness relaxes the constraints on what is considered correct and frees the evolutionary search process to make changes which affect the underlying structure of the web page HTML.
A screenshot only tells us if the page loaded as expected or not, and does not give us any gradient or measure of subsequent page functionality. A screenshot gives no way of telling apart a completely disfunctional page load from a partially functional version of the app. Some functionality is better than none, and Evolutionary algorithms rely on this gradient. In future work the HTML state could additionally be used to give partial functionality gradient.

\subsection{Operators}
We investigate the use of two operators, one which was developed by the authors of this paper, and the other, which was developed and made freely available by open source contributors. 
Both operators build an Abstract Syntax Tree (AST) of Javascript source files which are searched, manipulated and written back to a file. 

\subsubsection{Deletion Operator}
The simplest way to reduce page load time is to reduce the amount of computation done. To inspect this, we delete portions of the web app source code. Deleting code gives us some indication of the mutational robustness of the software we are targeting. If any portion of the code can be deleted without affecting the correctness of the resulting web page, then we can say that there is some level of redundancy in the code. Claims of robustness can only be made within the context of the operators and test cases used. If the correctness tests (pixel differences between screenshots) do not capture important features or functionality of the web app, then deletion may find performance improvements which turn out to have an undesirable effect on functionality.

A page load trace contains method calls but also which Javascript file the method calls were made from. We use this information to find where methods are called and defined within the Javascript source. 
We should also note that deletion is distinct from dead code removal. Only method names which appear in the page load trace are considered for deletion. As these methods appear in the trace, we know they have been executed and would not be removed by dead code removal.

\subsubsection{Loop Optimiser} 

The ``loop optimiser'' operator has made freely available\footnote{https://github.com/vihanb/babel-plugin-loop-optimizer} as a ``community'' contribution to the Babel project\footnote{https://babeljs.io/}. 
This source code transform is intended to improve the performance of loops. A potential issue with this type of operator is that it is not well tested\footnote{The version we used was specifically marked as experimental}. 
While community written program transforms can contain useful improvements, they are not written or tested with the same rigour that might be found in, for example, the transforms performed by a compiler. 

Similar community efforts to produce code transforms gives the Evolutionary Computation community a potential source of operators. Though an operator may perform a beneficial transform in most contexts, a developer will still need to test and validate the effect this transform has on their code. A potential impediment to the uptake of these transforms is that we do not have a high level of assurance that these transforms are rigorously semantics preserving. For languages that are dynamically typed, it is more difficult to ensure new transforms preserve semantics. 
If program transforms are to be crowd-sourced to some extent, we may end up with a scenario where the proliferation of transforms becomes an issue in itself for a developer as time is required to choose, apply and validate transforms. This motivates a Genetic Programming mutate-and-test approach to the application of community operators.

\subsection{Search Loop}
The search loop iterates through a list of file names from the web app and applies an operator to each one.
When applying deletion, the loop iterates through a list of method names and associated files. Method names and files are extracted from a page load trace and used to focus operators on statements and expressions which contain the method name. For each file, each mention of a method name in the AST is found. From each element in the AST which contains the method name, the parent statement or expression is deleted. The app is then loaded over HTTP via the chrome browser which captures a trace of the page load and a screenshot. Tracing will timeout after twice the time the oracle page load takes (3 seconds). Runtime and event counts are extracted from the trace (used in post-analysis only) and the screenshot is compared with the oracle screenshot to give correctness values. If the page is considered correct, the mutation is kept, otherwise the mutation is discarded and the next operator application is made. To be clear, the performance metrics are not used to guide the search process. The only metric which determines whether a change is kept, is whether the page loads as expected. The performance metrics are not used directly as part of the search loop but are used for post-analysis. When all possible operator applications have been made, we compare the most recent patch, which contains the culmination of all code changes, to the original. By virtue of the operator we inspect, if the page loads correctly after many deletions, we can perform more in-depth performance bench-marking to quantify the performance improvement found.

\section{Web Application}
\label{webapp}
In this paper we use an Angular2 web app which is written in typescript and is compiled to Javascript\footnote{Full web app code: https://github.com/mlaval/optimize-angular-app} It is this generated Javascript that we target for optimisation. The app is designed to contain redundancy and as an exemplar app with performance issues. The functionality provided by the app is to load a paragraph on the screen and display page load time\footnote{Demo: https://mlaval.github.io/optimize-angular-app/dev}. Page load takes under two seconds even though the resulting page would be far more effectively delivered as static HTML, instead of being loaded via Angular2 functionality. Any optimisations found will show a clearly measurable improvement in runtime as well as being easy to understand and validate.  In this scenario, Angular is far too heavyweight a solution for the resulting web page state. While this appears an easy optimisation target, it is however a challenging benchmark problem for an evolutionary approach and an excellent initial problem on which to validate our approach. The challenge lies in disentangling the extensive sequence of method calls which ultimately produce a static page of text. The oracle web application triggers over 5000 events to a depth of 242 nested method calls. Although Angular is designed to deliver more complex functionality, the increased level of abstraction nonetheless makes it difficult to understand interactions between many layers of sometimes anonymous and interleaved method calls for this simple app. Our hope is that an evolutionary algorithm utilising a range of operators which can reduce the dependencies and redundancies in this benchmark application will also be able to reduce other ``real world'' web applications in terms of load time.

\section{Results}
\label{results}

\begin{figure*}
  \includegraphics[width=\textwidth]{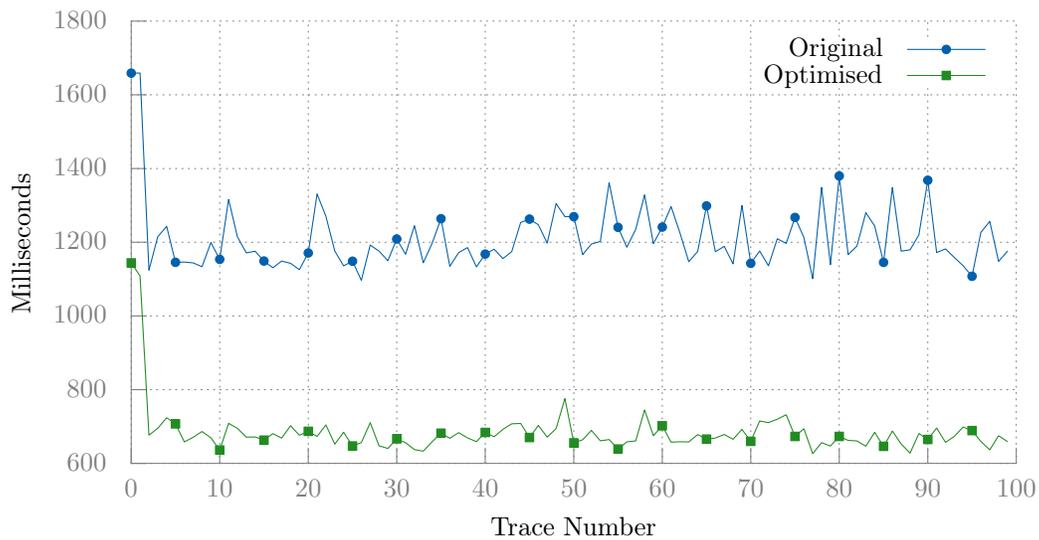}
  \vspace{-1.5cm}
  \caption{Trace of Original and Optimised web app versions}
  \label{fig:trace_warmup}
\end{figure*}

Figure~\ref{fig:trace_warmup} shows 100 repeated measurements of page load time for the original web app and an optimised version of the web app which was derived by the exhaustive application of the \textbf{deletion operator}. Each sequence of traces shows roughly the same pattern, though they appear offset by roughly 500ms. We notice that there is a warm-up factor when traces are repeatedly gathered on the same page as initial page load time measures are high in the first couple of traces when compared to the relatively stable measures gathered subsequently.

\begin{figure}
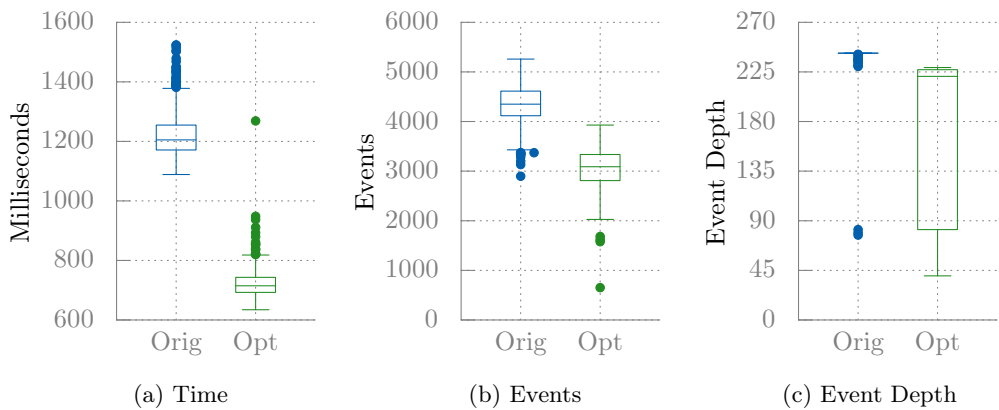

  \begin{subfigure}[t]{0.35\textwidth}
    \includegraphics[width=\linewidth]{time-box}
    \caption{Time}
    \label{fig:time_box}
  \end{subfigure}
  \hspace{-.6cm}
  \hfill
  \begin{subfigure}[t]{0.35\textwidth}
    \includegraphics[width=\linewidth]{events-box}
    \caption{Events}
    \label{fig:event_box}
  \end{subfigure}
  \hfill
  \hspace{-.6cm}
  \begin{subfigure}[t]{0.35\textwidth}
    \includegraphics[width=\linewidth]{eventdepth-box}
    \caption{Event Depth}
    \label{fig:eventdepth_box}
  \end{subfigure}
  \caption{Average Metrics (1000 samples) for Original (Orig) and Optimised (Opt) versions of the web app}
      \label{fig:time_event_depth}
\end{figure}

Figure~\ref{fig:time_event_depth} shows average values for time, number of events and largest event depth from 1000 samples taken after browser warm-up. On average, there is a 41\% saving in page load time. 30\% less events occurred and event depth was 26\% lower. Variance in measures was roughly the same except for event depth which varied significantly in the optimised version of the app. This may indicate that there are reduced dependencies in this version of the app which leave the browser more freedom in when and how events are triggered. Although a saving of 500ms is a lot in terms of page load time, the mean page load time for the optimised version of the app was still 720ms. Given that the final rendered page is mostly static and the page is loading from the local machine we should be able to achieve far lower load times with additional operators and further refinements of our approach leaving us ample room for future improvement on this benchmark problem.

22\% of deletions had no effect on functionality as measured
and we can say that the benchmark app we used is relatively robust to deletions. This is an interesting result as we expected removing code which appears in the load trace to be more destructive. 8966 lines were deleted out of a total of 17022 (52\%). 

Method names found in traces of the app appear 982 times in the app source code. For this particular benchmark app, as method calls were deleted, other method calls would appear in the trace. This is due to the redundancy contained in this benchmark application.

We found one iteration of our search loop, which performs parsing, AST traversal, file writing, page load over HTTPS, trace call tree building, call tree traversal and source code diffing took between 7 and 13 seconds. We feel this is not prohibitive for evolution in the browser, especially given that it is likely to be reduced on further refinement of our approach. 

We used the \textbf{Loop Optimiser} transform in our search loop but unfortunately no performance improvements were found. During experimentation we also applied the loop optimiser to all Javascript files in the web app and found that it resulted in a page that did not load. The existence and availability of such experimental operators justify a search-based approach which can discover what operators are applicable where. 

For GI research, these results reiterate the question as to what is the best way to go about operator design. We could have a wide range of specific operators which are only applicable to certain code patterns (loop optimisation), or have a few very general operators (delete, clone, replace) which are unlikely to leave a program in a fully functional state and only rarely improve performance.

\subsubsection{Limitations}
The main limitation of this work is that the benchmark app is very simple and contains a lot of redundancy by design. As a result, the findings give little indication as to how generalisable the approach is. When considering more real-world applications of this approach we feel it is unlikely that they will contain the same level of redundancy or be so cleanly structured that deleting lines of code will improve performance while leaving functionality similarly unaffected. 
Additionally, though comparing screenshots is enough to detect a ``correct'' page load for our benchmark app, a web app which provides user functionality would need extensive additional testing.

\section{Conclusion}
\label{conc}
We find that improvement of browser-based Javascript using iterative mutation and testing is possible. We reduce a benchmark app to only what code is necessary to provide the desired functionality, saving 41\% in runtime. We may compare code changes against those derived through program slicing techniques~\cite{ye2016efficient} and investigate if they produce similar performance improvements.
Finding performance improvements further to a minimal code ``slice'' will likely require more inventive operators. What operators to use is a pertinent open question. We highlight that community contributed code transforms may make ideal candidate operators for search-based approaches. 
Our findings also give a good base for expanding the work toward more complex ``real world'' applications with more detailed measures of functionality (e.g. HTML diff or navigation testing) and performance (e.g. user interface jitter). \\

\setlength{\parindent}{0pt} \small
\textbf{Acknowledgements}
This research is based upon works supported by Science Foundation Ireland under grant 13/IA/1850 and 13/RC/2094 which is co-funded under the European Regional Development Fund through the Southern \& Eastern Regional Operational Programme to Lero - the Irish Software Research Centre (\url{www.lero.ie}).

\bibliographystyle{splncs03}
\bibliography{cody-kenny}

\end{document}